\documentclass{ws-p8-50x6-00}

\def\PRC{{\em Phys. Rev.} C}

\def\bfi{\begin{figure}}
\def\efi{\end{figure}}
\def\fref#1{Fig.~\ref{#1}}
\def\eref#1{Eq.~(\ref{#1})}

\newcommand\de{\delta}
\newcommand\om{\omega}

\newcommand\q{\theta}

\newcommand\rp{r_\perp}

\begin{document}

\title{Probing the Space-Time evolution of Heavy Ion Collisions 
with Bremsstrahlung}

\author{J.I. Kapusta and S.M.H. Wong\footnote{presenter}}

\address{School of Physics and Astronomy, University of Minnesota,
Minneapolis, \\ MN 55455, U.S.A.
\\E-mail: kapusta@physics.spa.umn.edu 
\\E-mail: swong@nucth1.hep.umn.edu \ \ }


\maketitle

\abstracts{
We examine the bremsstrahlung photons emitted in the central collisions 
of two gold nuclei at the Relativistic Heavy Ion Collider. 
While the measurements of the final hadrons can reveal only the amount
of stopping, they tell us very little about the space-time evolution
of the bulk matter in the collisions. By using two extreme collision
scenarios, we argue that the low and medium energy bremsstrahlung photons
can, not only reveal the amount of stopping, but also the finer details 
of the space-time evolution of the charges.
}

\section{Introduction} 
\label{sec:intro}

Very soon at the Relativistic Heavy Ion Collider (RHIC) at Brookhaven
National Laboratory, experiments which could previously only be conducted at 
energies up to 9 GeV/nucleon can now be done at 100 GeV/nucleon in the
centre of mass frame. At these newly attainable energies, the probability for 
discovering new physics in the form of the formation of an entirely new 
many-body system under extreme conditions, the so-called quark-gluon plasma 
(QGP), should be greatly enhanced. There are many important issues that one 
should address. In this talk, we will focus on baryon stopping and the 
space-time evolution of the bulk charges in the system. 

The information on the degree of baryon stopping is important for
several reasons. The presence of net baryon density in the most energetic 
collision zone will affect the rates of particle production, that of 
parton chemical equilibration\cite{won}, etc. Theoretical estimates 
of the signatures of QGP, especially those that assume specific environmental 
conditions as the starting point, have the need for a reasonably 
good input of the net baryon concentration. The present method is to
measure the rapidity distributions of the hadrons. To do this, detectors 
must cover at least $2\pi$ of the collision centre. That means
no small number of hadronic detectors will have to be deployed in a
hemisphere. The rewards of such experimental efforts are somewhat 
limited because these measurements will only give us information about the 
surface of last interactions of the hadrons. They will not tell us how the 
hadrons actually get there. We will show that an easier method
exists for measuring stopping and that it can also provide information
on the space-time history of the net charge in the collisions.

\section{Measuring Charge Stopping with Bremsstrahlung}
\label{sec:brems}

As the nuclear matter collides in A-A collisions, QCD as the dominant 
interactions will slow down the relativistically moving nuclei. While
this is happening, the simple fact that accelerating charges must 
emit photons, the intensity of which depends on the accelerations
hence the degree of stopping, give us the possibility of measuring 
stopping by detecting photons instead. The first advantage is only
photons need to be detected. We shall see the second advantage shortly that 
the measurement is highly localized in spatial direction. These should 
simplify the detector design and construction. 

\bfi[t]
\epsfig{figure=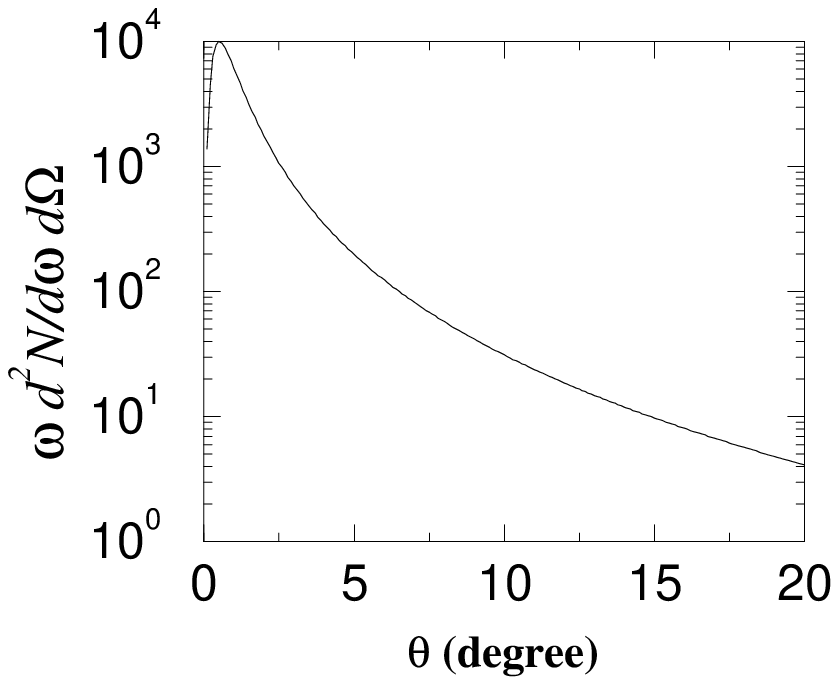,width=2.0in}
\hskip 1.50cm
\epsfig{figure=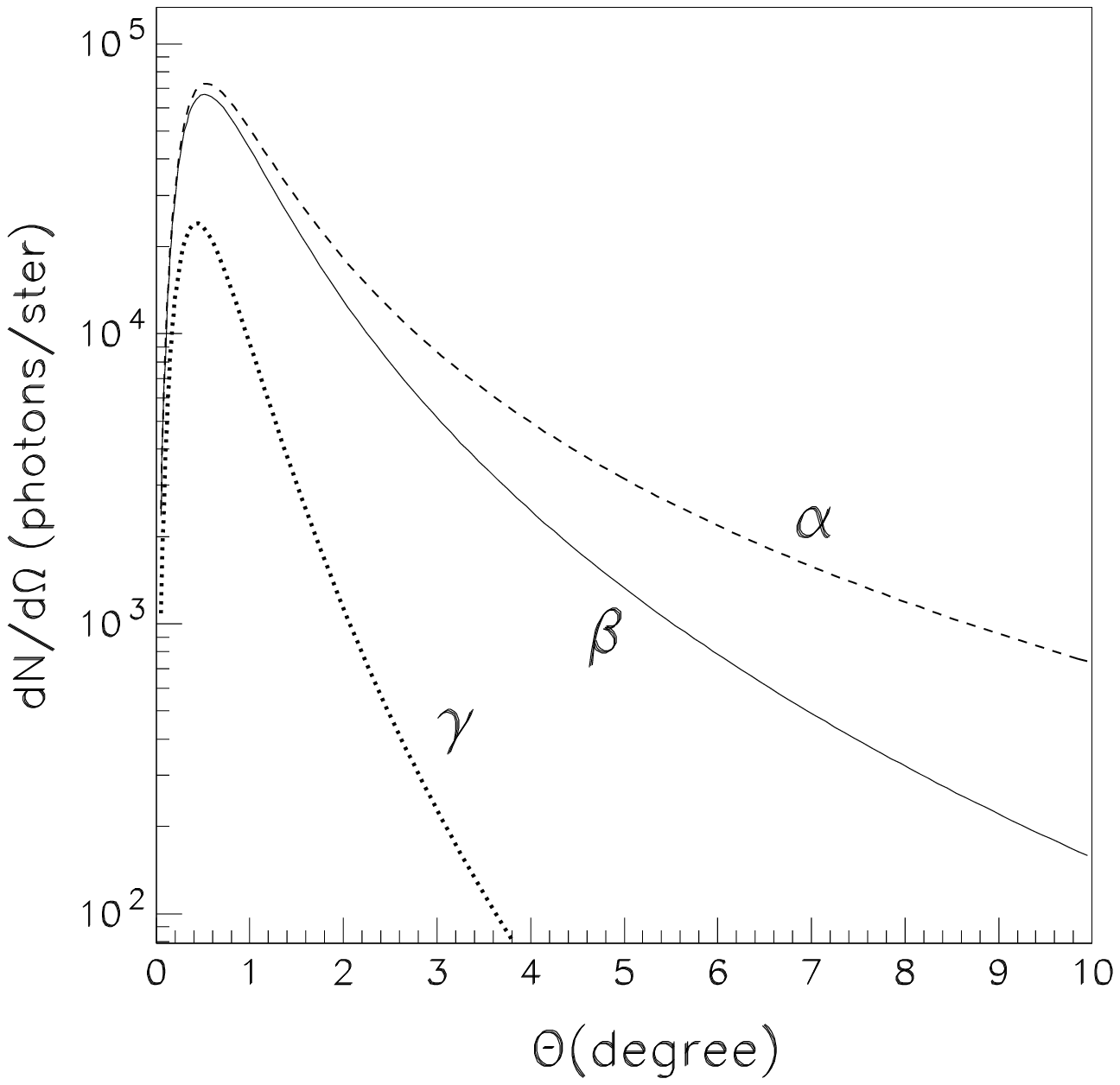,width=1.80in,height=1.60in}
\caption{Plot of the intensity of soft photons vs. $\q$ for a flat final
charge rapidity distribution (left). Plot of the number of soft photons 
vs. $\q$ when there is $(\alpha)$ full stopping, $(\beta)$ half transparency 
and $(\gamma)$ near complete transparency (right).}  
\label{f:or}
\efi

Using the standard formula\cite{jack} for the radiation of a time varying 
current, the initial highly Lorentz contracted nuclei traveling 
at $v_0 =\tanh y_0 \sim 1$ represented by the current 
$J_i (z,t) = v_0 \; \sigma \{ \de(z-v_0 t) - \de(z+v_0 t) \} \; \hat z \; ,
$
becomes 
$J_f (z,t) = \sigma \int d y \; \rho(y) v(y) \de (z-v(y) t) \; \hat z
$
when the charges have a final rapidity distribution $\rho(y)$. 
The intensity can be worked out to be 
\be \om
\frac{d^2 N}{d\omega d\Omega} = \frac{\alpha Z^2}{4\pi^2}
\sin^2\theta \left| F(\omega \sin\theta) \right|^2 \left|
\left[ \int dy \frac{v(y) \rho(y)}
{1-v(y)\cos\theta} - \frac{2v_0^2 \cos\theta}{1-v_0^2
\cos^2\theta} \right] \right| ^2 
\label{eq:form}
\ee
at an angle $\q$ from the beam or $z$-axis. For a flat
rapidity distribution, this is plotted in \fref{f:or}(left). 
The maximum intensity is at small angle close to the beam pipe. 
This is so because \eref{eq:form} is dominated at small angle 
by the second term when that denominator is small. This emission in the
extreme forward direction is of course a familiar relativistic effect. 
It is this effect that ensures only a few photon detectors need to be 
placed at precise locations for these measurements. 

Now to see if it is possible to use bremsstrahlung to distinguish between
the various degrees of stopping, all we need to do is to draw some $\rho(y)$
representing different cases and use these in \eref{eq:form} above. To be 
able to compare them, they must all give the same total multiplicity. This 
has been done in ref.\cite{jkcs} and \fref{f:or}(right) shows that this is 
indeed possible. Although the region of maximum intensity tends to be close
to the beam pipe, this plot shows that it is sufficient to point the photon 
detectors at a few degrees off the beam axis. So it is clear that by 
strategically placing a few detectors around the beam pipe, charge stopping 
can be measured by this much more localized detection method, whereas if 
hadrons are used for this purpose, many more detectors are required and 
they must cover a much wider area.

\section{LEXUS}
\label{sec:lexus}

The above was an illustration of an easier alternative to assessing
stopping via hadron rapidity distributions. We now turn
to the space-time evolution of the bulk charges in the collisions. 
To do this, we need a more realistic 2-D rapidity-transverse spatial
distribution. This can be readily generated from the LEXUS model\cite{lexus}
which was designed by choice to interpret A-A collisions as two clusters 
of nucleons undergoing rows on rows of multiple nucleon-nucleon scatterings. 
In this model, the parton language is deliberately not used at all. 
The philosophy behind this is that if the generated results cannot reproduce 
experimental data, then there must be more to the collisions than
nucleons colliding with each other. One could interpret QGP 
formation as an example of one such possibility. Since p-p collisions 
have essentially hyperbolic cosine distributions, as far as generating
nucleon rapidity distribution is concerned, there is only one parameter
which is the probability for a nucleon-nucleon collision to be hard
as opposed to diffractive or elastic. Once this is fixed from
p-p data, the rows on rows of nucleon collisions in Au+Au collisions
at RHIC energies can be performed to give the distribution plotted in 
Fig. 1 of ref.\cite{jkcs}. The shape of this distribution can be
captured in a parameterization
\be
\frac{d^3Q}{d^2r_{\perp}dy} = n_0 \frac{Z{\rm e}}{A} \frac{dl(r_{\perp},y)}{dy}
= n_0 \frac{Z{\rm e}}{A} \left[f(r_{\perp}) + g(r_{\perp})\cosh y \right] \, .
\label{eq:param}
\ee
The functions $f(\rp)$ and $g(\rp)$ can be found in ref.\cite{kap&won}.
Using this, the combined radiation amplitude from the target and
projectile can be written in terms of the acceleration $a(y,t)$ as
\bea A(\om,\q) &=& \sin\theta \frac{Z{\rm e}}{A} n_0 \int_0^R dr_{\perp} r_{\perp}
J_0(r_{\perp}\omega \sin\theta) \int_0^{y_0} dy \frac{dl(y,r_{\perp})}{dy}
\nonumber \\
&\times& \int_{-\infty}^{\infty} dt \exp\{i\omega \left(t - z(y,t)\cos\theta
\right)\} \frac{a(y,t)}{(1 -v(y,t) \cos\theta)^2} \nonumber \\
& & + \{(z,v,a) \longleftrightarrow (-z,-v,-a) \} \, .
\label{eq:pt_amp}
\eea
This paves the path for what are to follow in the next section.

\section{Two Scenarios of Space-Time Evolution}
\label{sec:2s}

As we mentioned in Sec. \ref{sec:intro}, measuring hadrons alone
will not tell us much about what happened to the bulk of the matter
in the intervening period after first nuclear contact but before the
final break up of the system. Below, we will show two collision scenarios 
which follow different space-time paths but both end up with the same final 
nucleon rapidity-transverse spatial distribution generated from LEXUS. 
This aspect guarantees that no measurement of the rapidity distribution 
of final state hadrons will be able to distinguish between them.  

The first scenario is Bjorken-like where the nucleons progressively
decelerate to the final rapidities. The duration of this deceleration
$t_f$ may or may not depend on how much nuclear matter each nucleon
has to transverse. The second scenario is Landau-like in which all
nucleons in the two initial nuclear clusters must come to a complete
stop instantaneously on initial nuclear contact. Subsequently, the
immense pressure existing in the highly excited and compressed region
accelerates and drives the nucleons to their final rapidities. 
The time $t_f$ required for the bulk matter to finally settle down must be
independent of where the individual nucleons or charges are located
in the system. The accelerations in both cases are expressed in the 
simple form
\be a(l,\rp) = \{ v_f(y)-v_0\} \; t^{-1}_f(\rp) 
             = \{\tanh y-\tanh y_0\} \; t^{-1}_f(\rp)  \; .
\ee
Inserting this into \eref{eq:pt_amp}, the bremsstrahlung radiation can
be worked out for each scenario.

\bfi[t]
\centerline{
\epsfig{figure=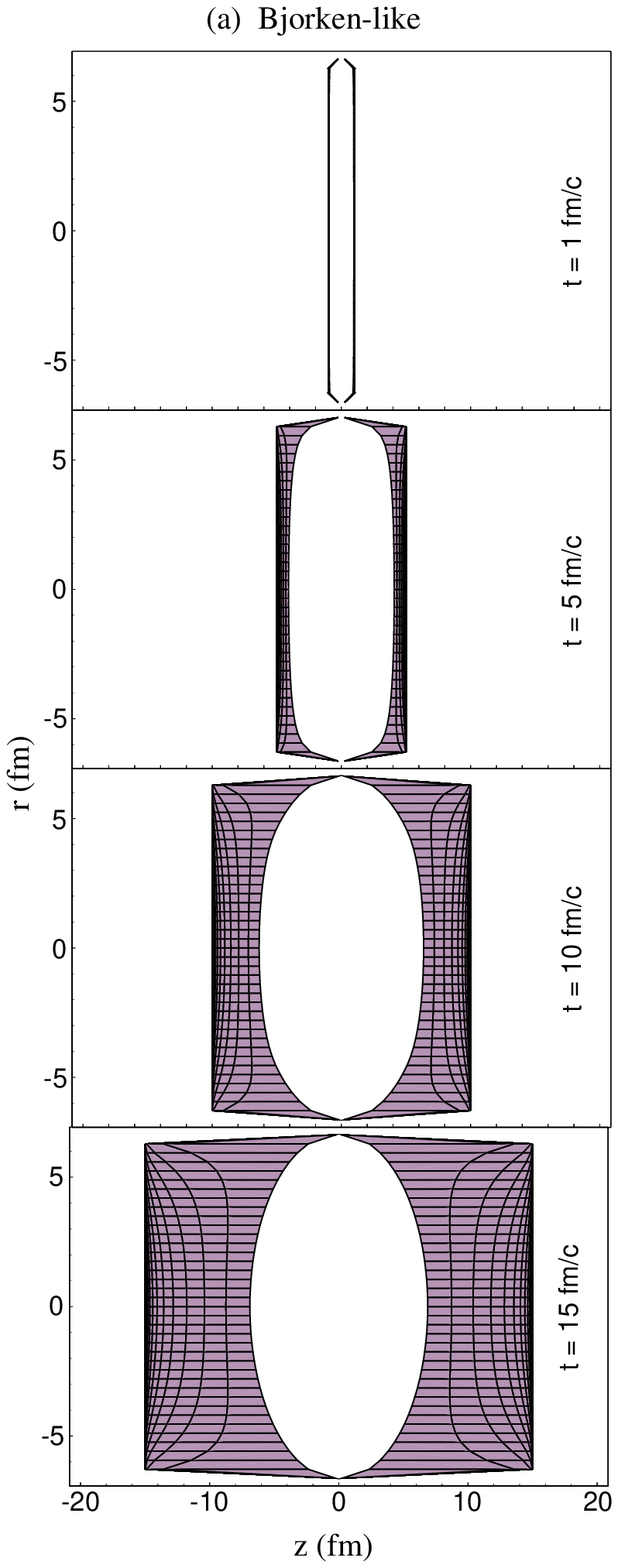,width=1.70in}
\hskip 1.0cm
\epsfig{figure=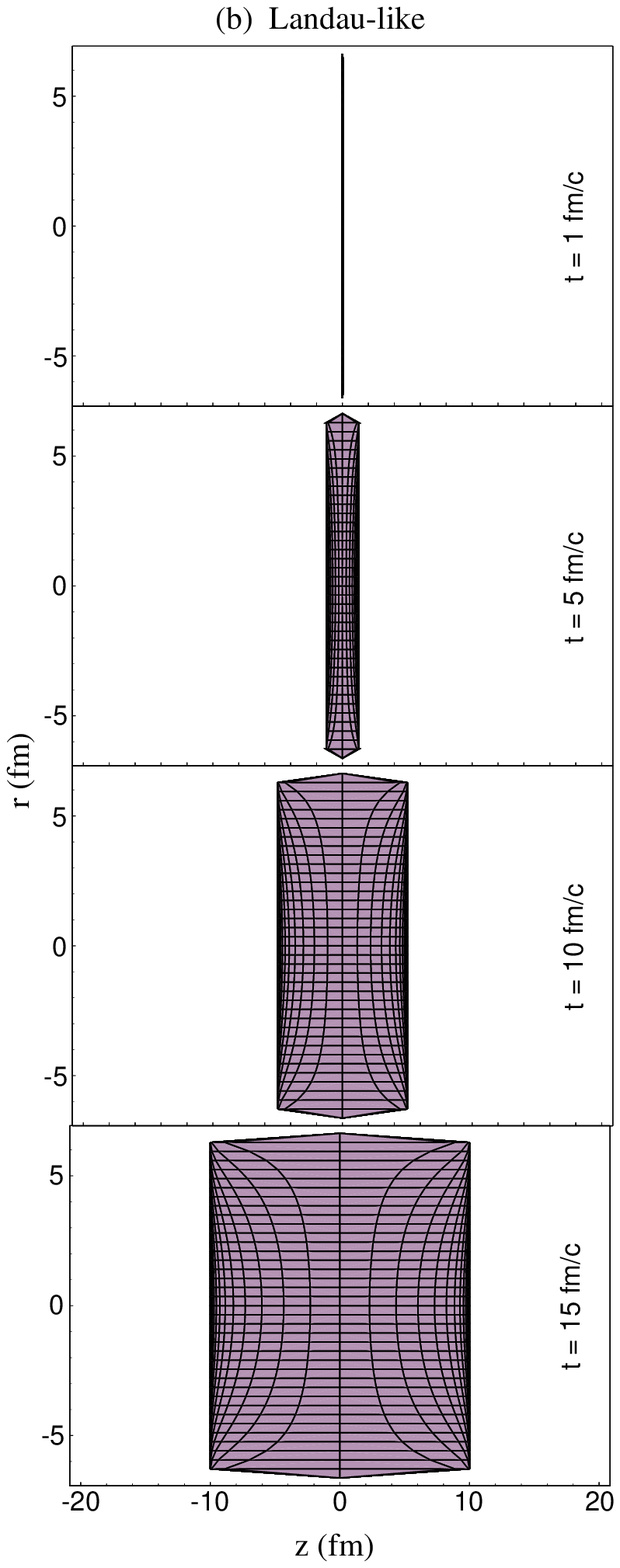,width=1.70in}
}
\caption{Two scenarios of the space-time evolution in a Au+Au collision.}
\label{f:sp_ti}
\efi

In \fref{f:sp_ti}, we show just one example of each scenario of 
comparable time scale. In this specific example, the duration for the 
deceleration in the Bjorken-like scenario depends on the amount of matter 
in the path of each nucleon, therefore there is a central oval-shaped 
charge free region. The Landau-like scenario on the other hand has spatially 
uniform acceleration and therefore there is no electrically neutral zone. 
In both cases, the highest densities and greatest variations in the 
concentration of the charges are at the front of the expanding matter as 
shown by the contours. The temporal-spatial distribution of the charges 
in the two scenarios are indeed very different even though the final rapidity 
distributions are the same. 

\bfi[t]
\centerline{
\epsfig{figure=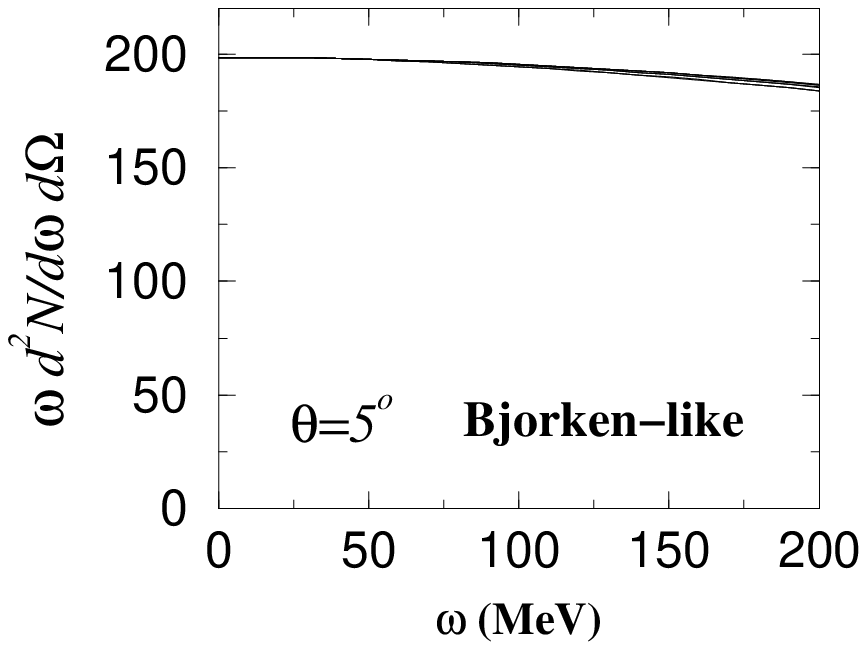,width=2.0in}
\hskip 1.0cm
\epsfig{figure=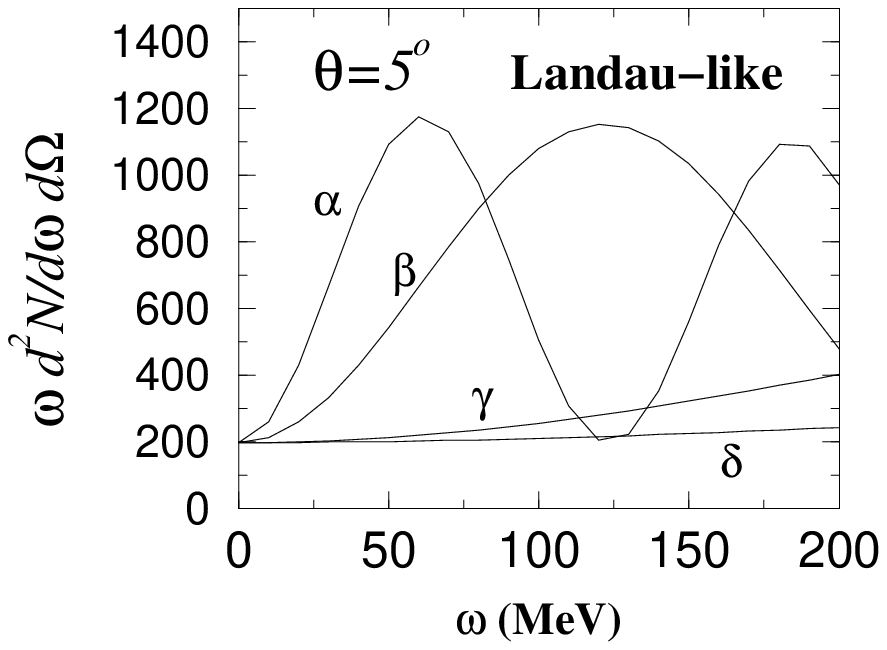,width=2.0in}
}
\caption{Whereas the bremsstrahlung spectra in the Bjorken-like scenario 
show almost no dependence on $t_f$ and very slowly variations with $\om$,
the Landau-like scenario shows oscillations and enhancements for
$t_f= 20\, (\alpha), 10\, (\beta), 2\, (\gamma)$ and $1\, (\de)$ fm/c.}
\label{f:l}
\efi

In \fref{f:l}, the variations of the intensities of the bremsstrahlung
emission with the photon energy $\om$ in each scenario are plotted at 
$\q = 5^o$ for various values of $t_f$. Whereas in the Bjorken-like
scenario, the intensity is very slowly varying with $\om$ and the 
variation of $t_f$ makes little difference, the Landau-like scenario
shows oscillations with $\om$ and the frequency clearly depends on $t_f$. 
Thus by examining the soft photon spectra, it is possible to tell one scenario 
from the other. In the second scenario, we have an added bonus of being able 
to measure the time scale for the bulk matter to settle down. The oscillations
are a result of interference due to the existence of two components
in the acceleration, the instantaneous stopping and the intense pressure 
driven expansion, of the bulk charges. The interference of the
two greatly enhances the intensities and manifests itself as oscillations
in the radiation spectrum.

\section*{Acknowledgments}
S.W. would like to thank the Organizing Committee for a very enjoyable 
and informative symposium. This work was supported by the U.S. Department 
of Energy under grant DE-FG02-87ER40328.

\end{document}